\documentclass[aps,prd,preprint,showpacs]{revtex4}

\usepackage{graphicx,amsmath,amsfonts,amssymb,mathrsfs,epic,eepic}

\allowdisplaybreaks[1]

\newcommand{\ket}[1]{\left| #1 \right\rangle}
\newcommand{\bra}[1]{\left\langle #1 \right|}

\newcommand{\bld}[1]{\boldsymbol{#1}}

\begin{document}

\title{Exact series solution to the two flavor neutrino oscillation problem
in matter}

\author{Mattias Blennow}\email{mbl@theophys.kth.se}
\affiliation{Division of Mathematical Physics, Department of Physics, Royal Institute of Technology (KTH), AlbaNova University Center, Roslagstullsbacken 11,
106 91 Stockholm, Sweden}

\author{Tommy Ohlsson}\email{tommy@theophys.kth.se}
\affiliation{Division of Mathematical Physics, Department of Physics, Royal Institute of Technology (KTH), AlbaNova University Center, Roslagstullsbacken 11,
106 91 Stockholm, Sweden}

\begin{abstract}
  In this paper, we present a real non-linear differential equation
  for the two flavor neutrino oscillation problem in matter with an
  arbitrary density profile. We also present an exact series solution
  to this non-linear differential equation. In addition, we
  investigate numerically the convergence of this solution for
  different matter density profiles such as constant and linear
  profiles as well as the Preliminary Reference Earth Model describing
  the Earth's matter density profile.  Finally, we discuss other
  methods used for solving the neutrino flavor evolution problem.
\end{abstract}


\maketitle

\section{Introduction}

In general, there are several phenomena and processes in physics, but
also in other fields of science such as chemistry, that can be
described in terms of a system with two (quantum mechanical) states
and a time-dependent Hamiltonian, \emph{i.e.}, so-called two-level
systems -- neutrino oscillations with two flavors being one such
system. Other representatives of such systems are, for example, a spin
1/2 particle in a time-dependent electromagnetic field having the
states `spin up' and `spin down', $K^0$-$\bar K^0$ mixing, a Josephson
device, nuclear magnetic resonance used for encoding bits of
information (\emph{i.e.}, quantum bits for a quantum computer), the
left and right chirality states of molecules in chemistry, \emph{etc.}
The problem of neutrino oscillations in matter, which we are concerned
with in this paper, is mathematically equivalent to a spin 1/2
particle in a magnetic field that is constant in one direction, zero
in another direction, and time-dependent in the last direction
\cite{Kim:1994dy}.

Neutrino oscillations have recently been extensively studied in the
literature
\cite{Cleveland:1998nv,Fukuda:1998mi,Apollonio:1999ae,Apollonio:2002gd,Ahn:2002up}
and they act as the most plausible description of both the solar
\cite{Cleveland:1998nv} and atmospheric \cite{Fukuda:1998mi} neutrino
problems. At an early stage, neutrino oscillations were mainly
investigated with two flavors and without including matter
effects. Nowadays, we know that there are at least three neutrino
flavors and that matter effects are important. For example, in matter,
the so-called Mikheyev--Smirnov--Wolfenstein (MSW) effect
\cite{Wolfenstein:1978ue,Mikheev:1985gs} can take place, which is an
amplifying resonant effect due to the presence of matter. However, in
most situations, neutrino oscillations can be effectively investigated
with two flavors, since the leptonic mixing in the 1-3 sector is
indeed small \cite{Apollonio:1999ae,Apollonio:2002gd} leading to the
fact that the full three flavor scenario can be decoupled into two
effective two flavor scenarios, each of which can be studied
separately.

In this paper, we present an exact analytic solution to the two flavor
neutrino oscillation problem in matter.  Since there are many similar
two-level systems, as discussed above, our solution will also be
interesting and applicable to this kind of systems. However, before we
proceed to present our solution, we will give a brief overview of what
has previously been done in this field. Note that this overview is not
presented in chronological order. First, in
Refs.~\cite{Ohlsson:1999um,Akhmedov:1988kd}, the neutrino flavor
evolution has been investigated by a discretization of the effective
potential. Second, exact solutions exist for a number of specific
effective potentials
\cite{Wolfenstein:1978ue,Barger:1980tf,Lehmann:2000ey,Osland:1999et}.
Third, in Refs.~\cite{Smirnov:1987mk,Balantekin:1988aq}, the evolution
was studied by using an adiabatic approximation. Fourth, approximate
solutions valid for small effective potentials have recently been
studied in detail \cite{Ioannisian:2004jk}. Finally, there have also
been other attempts to write the evolution in terms of a second order
non-linear ordinary differential equation \cite{Fishbane:2000dc}.
However, this has been done for the neutrino oscillation probability
amplitudes and not for the neutrino oscillation probabilities. The
advantage of working with a non-linear differential equation for the
oscillation probability rather than a linear system of differential
equations for the probability amplitudes is that we only have one real
variable instead of two complex variables. The disadvantage is that
the resulting differential equation is non-linear.

This paper is organized as follows. In Sec.~\ref{sec:evol}, the
neutrino flavor evolution in matter with two flavors is studied and a
second order non-linear ordinary differential equation for the
neutrino oscillation probability is derived. Then, in
Sec.~\ref{sec:expansion}, we perform series expansions of both the
neutrino oscillation probability and the effective potential in order
to solve the differential equation presented in
Sec.~\ref{sec:evol}. Next, in Sec.~\ref{sec:convergence}, we continue
by studying the numerical convergence of the solution for a number of
different effective potentials and baselines. In
Sec.~\ref{sec:methods}, we present a brief summary of other methods
for solving the neutrino evolution in matter. Finally, in
Sec.~\ref{sec:summary}, we summarize our results and give our
conclusions.

\section{Neutrino flavor evolution in matter}
\label{sec:evol}

When neutrinos propagate in matter, neutrino flavors are affected
differently by coherent forward scattering against the matter
constituents. Assuming that there are no sterile neutrinos, the effect
of matter is to add an effective potential to $\nu_e$, this effective
potential is given by $V(t) = \sqrt 2 G_F N_e(t)$, where $G_F$ is the
Fermi coupling constant and $N_e(t)$ is the electron number density.

In the two flavor case, the time evolution of a neutrino state
$\ket{\nu(t)} = (\ket{\nu_e(t)} \ \ket{\nu_x(t)})^T$ is given by
\begin{equation}
{\rm i} \frac{{\rm d}\ket{\nu(t)}}{{\rm d}t} = (H_{\rm vac} + H_{\rm mat}(t)) \ket{\nu(t)},
\end{equation}
where $H_{\rm vac} = U {\rm diag}(m_1^2, m_2^2) U^\dag / 2E$ is the
free Hamiltonian in vacuum, $H_{\rm mat}(t) = {\rm diag}(V(t), 0)$ is
the addition to the free Hamiltonian due to matter effects, and
\begin{equation}
U = \left(\begin{array}{cc} c & s \\ -s & c\end{array}\right) \ 
\end{equation}
is the leptonic mixing matrix in vacuum. Here $c \equiv \cos\theta$,
$s \equiv \sin \theta$, and $\theta$ is the leptonic mixing
angle. Adding or subtracting terms proportional to the unity operator
to the total Hamiltonian $H(t) = H_{\rm vac} + H_{\rm mat}(t)$ will
only contribute with an overall phase to the neutrino state
$\ket{\nu(t)}$, and thus, does not affect the neutrino oscillation
probabilities. Using this fact, the total Hamiltonian may be written
as
\begin{eqnarray}
H(t) &=& \frac 12 \left(\begin{array}{cc}
V(t)-\frac{\Delta m^2}{2E}\cos 2\theta & \frac{\Delta m^2}{2E}\sin2\theta \\
\frac{\Delta m^2}{2E}\sin 2\theta & \frac{\Delta m^2}{2E}\cos 2\theta-V(t)
\end{array}\right) \nonumber \\
&=& \frac 12 \left[
\sigma_1 \frac{\Delta m^2}{2E}\sin2\theta +
\sigma_3 \left(V(t)-\frac{\Delta m^2}{2E}\cos 2\theta\right)\right],
\end{eqnarray}
where the $\sigma_i$'s ($i = 1,2,3$) are the Pauli matrices and
$\Delta m^2 \equiv m_2^2 - m_1^2$ is the mass squared difference
between the two mass eigenstates in vacuum.

The density matrix $\rho(t) = \ket{\nu(t)}\bra{\nu(t)}$ can be
parameterized as $\rho(t) = (\bld 1 + \bld S(t) \cdot \bld \sigma)/2$,
where $\bld 1$ is the unity matrix, $\bld \sigma = (\sigma_1 \
\sigma_2 \ \sigma_3)^T$ is the vector of Pauli matrices, and $\bld
S(t)$ is a vector such that $\bld S(t)^2 = 1$. Differentiating the
density matrix $\rho$ with respect to time $t$, the equation of motion
for $\bld S(t)$ becomes
\begin{equation}
\dot{\bld S}(t) = \bld S(t) \times \bld B(t),
\end{equation}
where $\bld B(t) \equiv g \bld e_1 + f(t) \bld e_3$ and we have
defined $g \equiv -\sin(2\theta) \Delta m^2 / 2E$ and $f(t) \equiv
\cos(2\theta) \Delta m^2 / 2E - V(t)$. Note that $g$ is independent of
time $t$. The probability of neutrinos produced as $\nu_e$ to
oscillate into $\nu_x$ (where $\nu_x$ is some linear combination of
$\nu_\mu$ and $\nu_\tau$) is now given by $P(\nu_e \rightarrow \nu_x)
\equiv P_{ex} = (1-S_3(t))/2$. With the parameterization
\begin{equation}
\bld S(t) \equiv \left(\begin{array}{c}
\sin \alpha \cos \beta \\
\sin \alpha \sin \beta \\
\cos \alpha
\end{array}\right),
\end{equation}
where $\alpha = \alpha(t)$ and $\beta = \beta(t)$, we obtain the
following non-linear system of ordinary differential equations
\begin{eqnarray}
\dot \beta &=& g \cot \alpha \cos \beta - f, \\
\dot \alpha &=& g \sin \beta .
\end{eqnarray}
Eliminating $\beta$ from the above expressions, we obtain the
differential equation
\begin{equation}
\label{eq:dealpha}
[\ddot \alpha + \cot \alpha (\dot \alpha^2 - G)]^2 =
F(t) (G - \dot \alpha^2),
\end{equation}
where $F(t) \equiv f(t)^2$ and $G \equiv g^2$.

Now, we make the substitution $p = S_3(t) = \cos \alpha$ after which
Eq.~(\ref{eq:dealpha}) becomes
\begin{equation}
\label{eq:dep}
(\ddot p + G p)^2 = F(t) [G(1-p^2)-\dot p^2].
\end{equation}
Note that $F = 0$ corresponds to the so-called MSW resonance condition
$\cos(2\theta)\Delta m^2/2E = V$. In this case, Eq.~(\ref{eq:dep})
takes the simple form
\begin{equation}
\ddot p + G p = 0
\end{equation}
with the trivial solutions $p = A \cos(gt) + B \sin(gt)$ just as
expected.

In general, the expression for $P_{ex}$ is known for constant matter
density and is given by \cite{Wolfenstein:1978ue}
\begin{equation}
\label{eq:constsol}
P_{ex} = \sin^2(2\tilde\theta) \sin^2\left(\frac{\Delta \tilde m^2}{4E}t\right)
= \frac{G}{F+G} \sin^2\left(\frac{\sqrt{F+G}}{2}t\right),
\end{equation}
where $\tilde\theta$ is the effective leptonic mixing angle in matter
and $\Delta \tilde m^2$ is the effective mass squared difference in
matter. Using that $p = 1 - 2P_{ex}$, it is a matter of trivial
computation to show that this is the solution to Eq.~(\ref{eq:dep})
with constant $F$, which corresponds to any constant matter density.

In the three ($n$) flavor case, the density matrix can be parameterized
by four [$2(n-1)$] real parameters. If we would adopt our approach to
the three ($n$) flavor case, then we would end up with a system of
seven [$2(n-1)$] non-linear ordinary differential equations, which, in
principle, can be solved in a manner analogous to the one described
above for the two flavor case.

\section{Series expansion of the solution}
\label{sec:expansion}

In order to solve the propagation of neutrinos in matter with
arbitrary density profiles, we adopt the method of series
expansion. We suppose that neutrinos are produced as $\nu_e$ and then
propagate through a given effective potential $V(t)$, this gives the
initial values $p(0) = 1$ and $\dot p(0) = 0$. Series expanding the
effective potential $V(t)$ and the quantity $p(t)$, we obtain the
following expressions
\begin{eqnarray}
V(t) &=& \sum_{n=0}^\infty V_n t^n, \\
p(t) &=& \sum_{n=0}^\infty p_n t^n,
\end{eqnarray}
where the coefficients $V_n$ ($n = 0,1,\ldots$) define the effective
potential and where we wish to compute the coefficients $p_n$ ($n =
0,1,\ldots$). By using the relation between $f$ and $V$, we obtain
\begin{eqnarray}
f(t) &=& \sum_{n=0}^\infty f_n t^n, \quad
f_n = \delta_{n0} \frac{\Delta m^2}{2E}\cos(2\theta)-V_n, \\
F(t) &=& \sum_{n=0}^\infty F_n t^n, \quad
F_n = \sum_{k=0}^n f_k f_{n-k}.
\end{eqnarray}

Inserting the above expressions into Eq.~(\ref{eq:dep}) and
identifying terms of the same order in $t$ gives the relation
\begin{eqnarray}
F_n G &=& \sum_{s=0}^n
(s+2)(s+1)(n-s+2)(n-s+1)p_{s+2}p_{n-s+2}+
\nonumber \\
&&\sum_{s=0}^n \left[ 2G(s+2)(s+1)p_{s+2}p_{n-s} + G^2p_s p_{n-s}\right] +
\nonumber \\ 
&& \sum_{s=0}^n F_{n-s}\sum_{k=0}^s
G p_k p_{s-k} + (k+1)(s-k+1)p_{k+1}p_{s-k+1}. \label{eq:pneq}
\end{eqnarray}
For $n=0$ with the given initial conditions, Eq.~(\ref{eq:pneq}) is a
second order equation in $p_2$ with $p_2 = -G/2$ as a double
root. This corresponds well to the fact that at $t=0$, the right-hand
side of Eq.~(\ref{eq:dep}) vanishes for the given initial conditions
and we are left with the equation $\ddot p(0) = -Gp(0)$. For $n=1$,
Eq.~(\ref{eq:pneq}) is trivially fulfilled (given the assumed initial
conditions, terms with $p_{n+2}$ will appear with the prefactor
$Gp_0+2p_2$ only), while the solution to the equation for $n=2$ is
simply $p_3 = 0$.

Also the solution for $n=3$ is now trivially fulfilled, since the terms
including $p_{n+1}$ also cancel for $n \geq 3$. For $n=4$, the
equation is a second order equation in $p_4$ with the solutions
\begin{equation}
p_4 = \frac{G^2}{24}\quad {\rm and} \quad p_4 = \frac{G(G+F_0)}{24}.
\end{equation}
Of these two solutions, only the latter will be a solution to our
problem, this is easily checked by inserting the known solution in the
case of constant effective potential from Eq.~(\ref{eq:constsol}).

For $n \geq 5$, Eq.~(\ref{eq:pneq}) is now linear in $p_n$. For $n
\geq 6$, we obtain a solution for $p_n$ in terms of lower order $p_k$,
$G$, and $F_s$, where $k < n$ and $s \leq n-4$. This expression is the
following recurrence relation
\begin{eqnarray}
p_n &=& - \frac{1}{G(n^2-3n+2)F_0} \left[
F_1 \sum_{s=1}^{n-2} (s+1)(n-s)p_{s+1}p_{n-s}\right. + \nonumber \\
&&
G(G+F_0)\sum_{s=2}^{n-2}p_s p_{n-s} + 
F_0 \sum_{s=3}^{n-3}(s+1)(n-s+1)p_{s+1}p_{n-s+1} + \nonumber \\
&&
2G \sum_{s=2}^{n-4}(s+2)(s+1)p_{n-s}p_{s+2} +
G \sum_{s=4}^{n-1}F_{n-s} \sum_{k=0}^s p_k p_{s-k} + \nonumber \\
&&
\sum_{s=4}^{n-2} F_{n-s} \sum_{k=0}^s (k+1)(s-k+1)p_{k+1}p_{s-k+1} + 
\nonumber \\
&&
\left. \sum_{s=3}^{n-3}(n-s+2)(n-s+1)(s+2)(s+1)p_{n-s+2}p_{s+2} \right].
\label{eq:pn}
\end{eqnarray}
For the first few coefficients we obtain
\begin{eqnarray}
p_0 &=& 1, \nonumber \\
p_1 &=& 0, \nonumber \\
p_2 &=& -\frac G2, \nonumber \\
p_3 &=& 0, \nonumber \\
p_4 &=& \frac{G}{24}(G+F_0), \nonumber \\
p_5 &=& \frac{GF_1}{48}, \nonumber \\
p_6 &=& -\frac{G(4G^2 F_0 + 8G F_0^2 + 4F_0^3 +F_1^2 - 
36 F_0 F_2)}{2880 F_0}, \nonumber \\
p_7 &=& -\frac{G(8G F_0^2 F_1 + 8F_0^3 F_1 - F_1^3 + 
4 F_0 F_1 F_2 - 48 F_0^2F_3)}{5760 F_0^2}, \nonumber \\
p_8 &=& \frac{G}{645120 F_0^3}\{
48GF_0^5 + 16F_0^6 - 63 F_1^4 + 16F_0^4(3G^2 - 34F_2) + \nonumber \\
&&
312 F_0 F_1^2 F_2 + 8F_0 (GF_1^2 - 30F_2^2 - 48F_1 F_3) + \nonumber \\
&&
4F_0^3 [4(G^3-34G F_2 +240F_4) - 53F_1]\}.
\label{eq:terms}
\end{eqnarray}
As can be observed by setting $F_k = 0$, the solution for $F = 0$,
\emph{i.e.}, at the MSW resonance, is just the series expansion for $p
= \cos(gt)$, which is clearly as expected.

\section{Convergence of the solution}
\label{sec:convergence}

In order to test our solution, we perform a number of numerical
tests. First of all, we give an overview of how we approximate the
electron number density (\emph{i.e.}, in principle, the effective
potential) by a polynomial. Then, we proceed by confirming that our
solution really converges nicely towards the simple trigonometric
function that is the exact solution for a constant electron number
density. In this case, we also study the convergence of the energy
dependence of the neutrino oscillation probability $P_{ex}(L)$ for a
baseline of $L = 3000$ km. After the constant electron number density
case, we investigate the case of a linear effective potential, and
finally, we study the case of the Preliminary Reference Earth Model
(PREM) \cite{Dziewonski:1981xy}.

In the numerical calculations, we have used the mixing angle $\theta =
13^\circ$ and the mass squared difference $\Delta m^2 = 2\times
10^{-3}$ eV$^2$ \cite{SK}. The value of $\theta$ approximately
corresponds to the upper limit on the leptonic mixing angle
$\theta_{13}$ from the CHOOZ experiment with $\Delta m^2 = 2\times
10^{-3}$ eV$^2$ \cite{Apollonio:2002gd}. The reason to use this
particular choice of parameters is that $\theta_{13}$ and the large
mass squared difference give the main effects to neutrino oscillations
from $\nu_e$ into other flavors for the baselines and energies we have
studied (for $L = 3000$ km and $E = 1$ GeV, the neutrino oscillations
governed by the small mass squared difference contribute with an
approximate addition of $0.05$ to the neutrino oscillation probability
$P_{ex}$, for shorter baselines and higher energies, this effect
decreases), see for example Refs.~\cite{Bilenky:1987ty}. The reason
for using the upper bound value for the mixing angle $\theta$ and not
some smaller value is that we wish to study the behavior of our
solution rather than to make any precise predictions about the
neutrino oscillation probabilities.

\subsection{Series expansion of the effective potential}

In order to use the series solution, which was obtained in the
previous section, we will need the coefficients $V_n$. In general, for
a given baseline length $L$, the effective potential $V(t)$ can be
expanded in terms of Legendre polynomials, \emph{i.e.},
\begin{equation}
V(x) = \sum_{n=0}^\infty c_n P_n\left(x\right), 
\quad P_n(x) = \frac{1}{n! 2^n} \frac{{\rm d}^n}{{\rm d}x^n}[(x^2-1)^n], \quad
c_n = \frac{2n+1}{2} \int_{-1}^1 V(x)P_n(x) {\rm d}x,
\end{equation}
where $x \equiv 2t/L-1$. For numerical treatments, we cannot use the
entire expansion in Legendre polynomials because of finite computer
memory and finite computer time. However, if we assume that the
coefficients $c_n$ are negligible for $n > N$, where $N$ is some
integer, then we have a polynomial approximation
\begin{equation}
V(x) \simeq \sum_{n=0}^N c_n P_n\left(x\right)
\end{equation}
of the effective potential. Clearly, given any polynomial $V(t)$, it
is a trivial matter to extract the coefficients $V_n$. This approach
turns out to be quite handy in the case of the PREM profile, which is
discussed below.

\subsection{Constant matter density}

\begin{figure}
\begin{center}
\includegraphics[width=12cm]{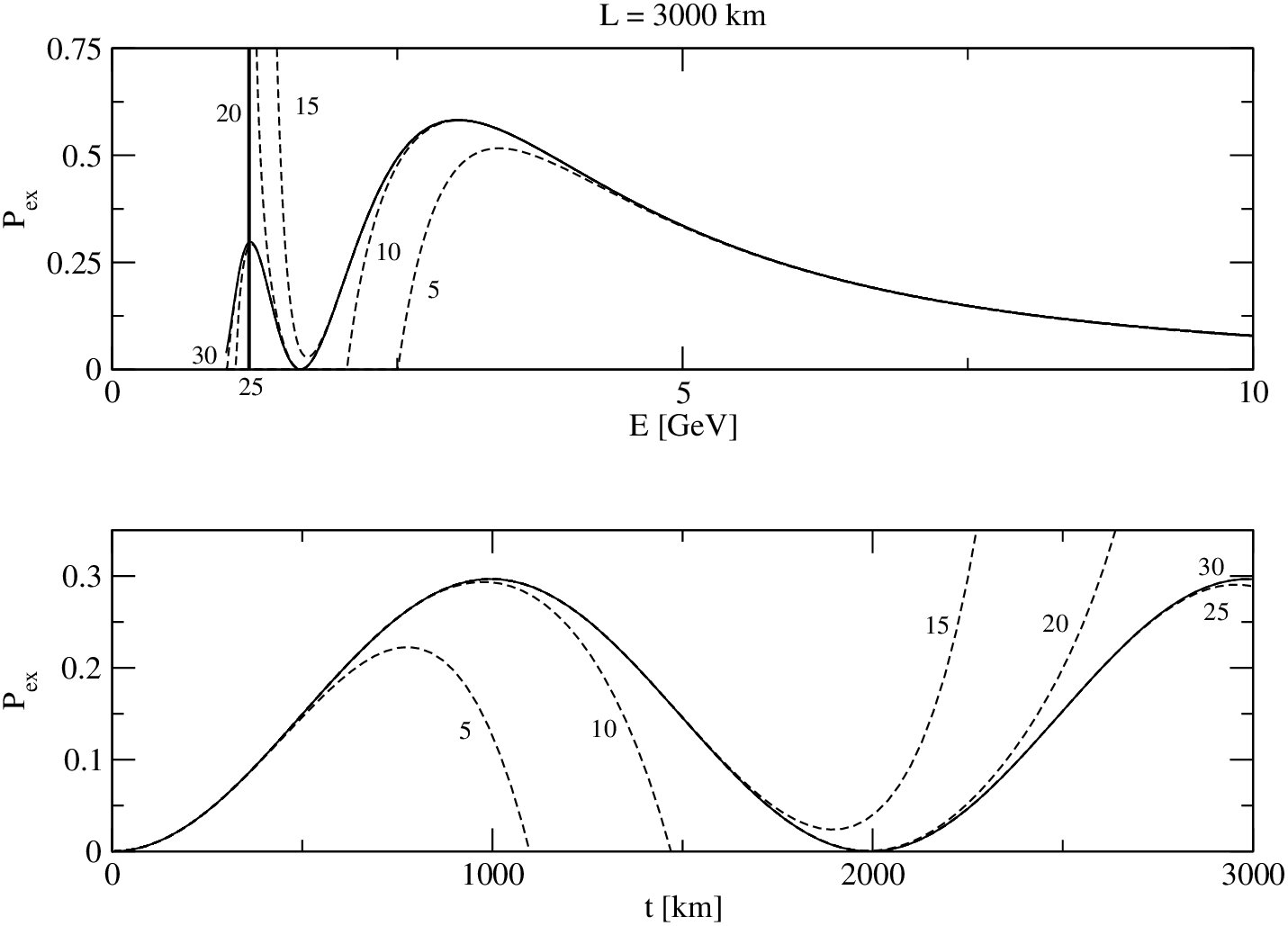}
\caption{The neutrino oscillation probability $P_{ex}$ as a function
  of energy and time, respectively. Upper panel: The convergence of
  the energy spectrum given by our series expansion for a constant
  electron number density profile with $N_e = N_{e,\rm core}/3$. Lower
  panel: The convergence of the series expansion for $E = 1.2$ GeV,
  corresponding to the bold line in the energy spectrum. The solid
  curves correspond to the exact solutions and the dashed curves
  correspond to the series expansion. The numbers correspond to the
  number of terms used in the series expansion. The solid vertical
  line in the upper panel corresponds to the energy used for the lower
  panel.}
\label{fig:const3000}
\end{center}
\end{figure}

The first case we study numerically is the case of a constant
effective potential. We use the baseline length $L = 3000$ km and the
electron number density $N_e = N_{e,\rm core}/3$, where $V_{\rm core}
= \sqrt{2} G_F N_{e,\rm core} \simeq 5.6 \times 10^{-19}$ MeV
corresponds to a matter density of about $13$ g/cm$^3$, which is the
maximum matter density in the Earth's core
\cite{Dziewonski:1981xy}. In this case, the coefficients $V_n$ are
easily obtained as $V_0 = V(t)$ and $V_n = 0$ for $n > 0$. Since the
exact solution to this problem is known \cite{Wolfenstein:1978ue}, we
focus on the convergence of our solution for $P_{ex}$, both in the
energy spectrum and the time evolution. The numerical results are
shown in Fig.~\ref{fig:const3000}.
In this figure, we can observe that approximately 20 terms are needed to
reconstruct one period of oscillation and that the convergence is
indeed the same as for a simple trigonometric function.

\subsection{Linearly varying matter density}

\begin{figure}
\begin{center}
\includegraphics[width=12cm]{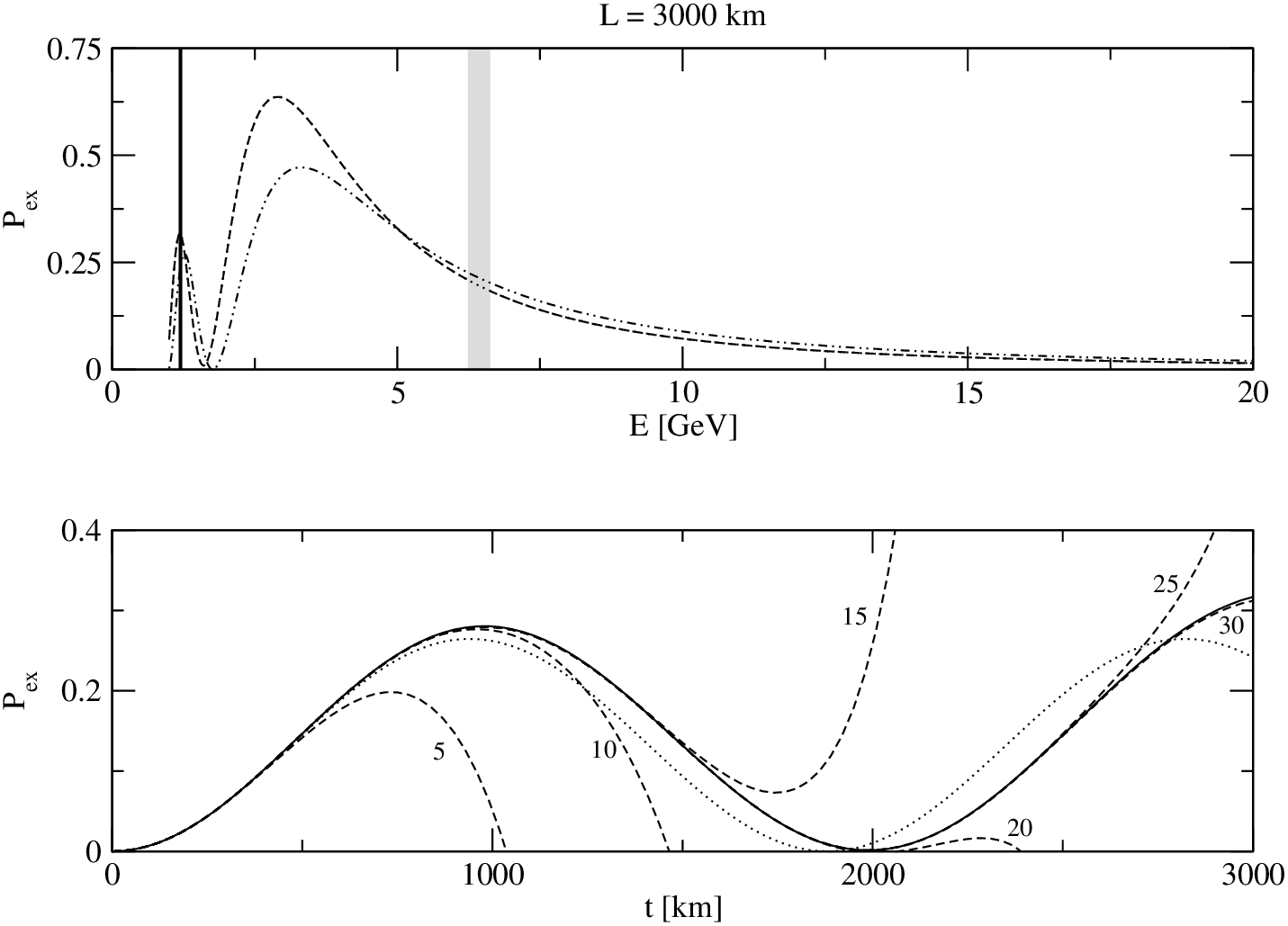}
\caption{The energy spectrum of the neutrino oscillation probability
  $P_{ex}$ for a linear profile (upper panel) with $N_e = N_{e,\rm
    core}(1+t/L)/4$ along with the convergence of the solution for $E
  = 1.2$ GeV (lower panel).  In the energy spectra, the dotted curve
  corresponds to the numerical solution for the given profile, the
  dashed curve corresponds to our series solution, where we have
  included the first 35 terms, and the dash-dotted curve corresponds
  to an approximation of constant matter density. The solid vertical
  line corresponds to the energy used in the lower panel and the series
  solution is not plotted in the shaded region where it breaks down
  numerically. In the time evolution plot, the solid curve corresponds
  to the numerical solution, the dotted curve corresponds to the
  approximation of constant matter density, the dashed curves
  correspond to our series solution for different numbers of included
  terms, and the numbers correspond to the number of terms used for
  each of these curves.}
\label{fig:linear3000}
\end{center}
\end{figure}

Now, we turn our interest towards the case of a linearly varying
effective potential. In particular, we study a baseline of $L = 3000$
km, where the electron number density is given by $N_e(t) = N_{e,\rm
core}(t/L + 1)/4$. As in the case of constant effective potential, it
is again easy to obtain the coefficients $V_n$ from our equation for
$V(t)$. Performing the numerical calculations result in
Fig.~\ref{fig:linear3000}.
In this figure, we have excluded the plot for our solution in the
shaded region, which roughly corresponds to an energy equal to the
resonance energy of the effective potential $V = V_0$, where the
solution breaks down numerically. The reason for this breakdown can be
found in Eq.~(\ref{eq:pn}), where we repeatedly divide by $F_0$. For
the resonance energy corresponding to $V = V_0$, we have $F_0 \sim 0$,
which leads to large absolute values of numbers that should add up to
a number between zero and one. Due to finite machine precision, we have
numerical errors as a result.

Apart from neutrino energies near the resonance energy, we can observe
that we again obtain a nice convergence of both the energy spectrum
and the time evolution, where we reproduce one full oscillation by
approximately 20 terms of our series expansion. It should be pointed
out that this case of linearly varying matter density has no known
application to experiments and only serves as an illustrative example.

\subsection{PREM profile}

\begin{figure}
\begin{center}
\includegraphics[width=12cm]{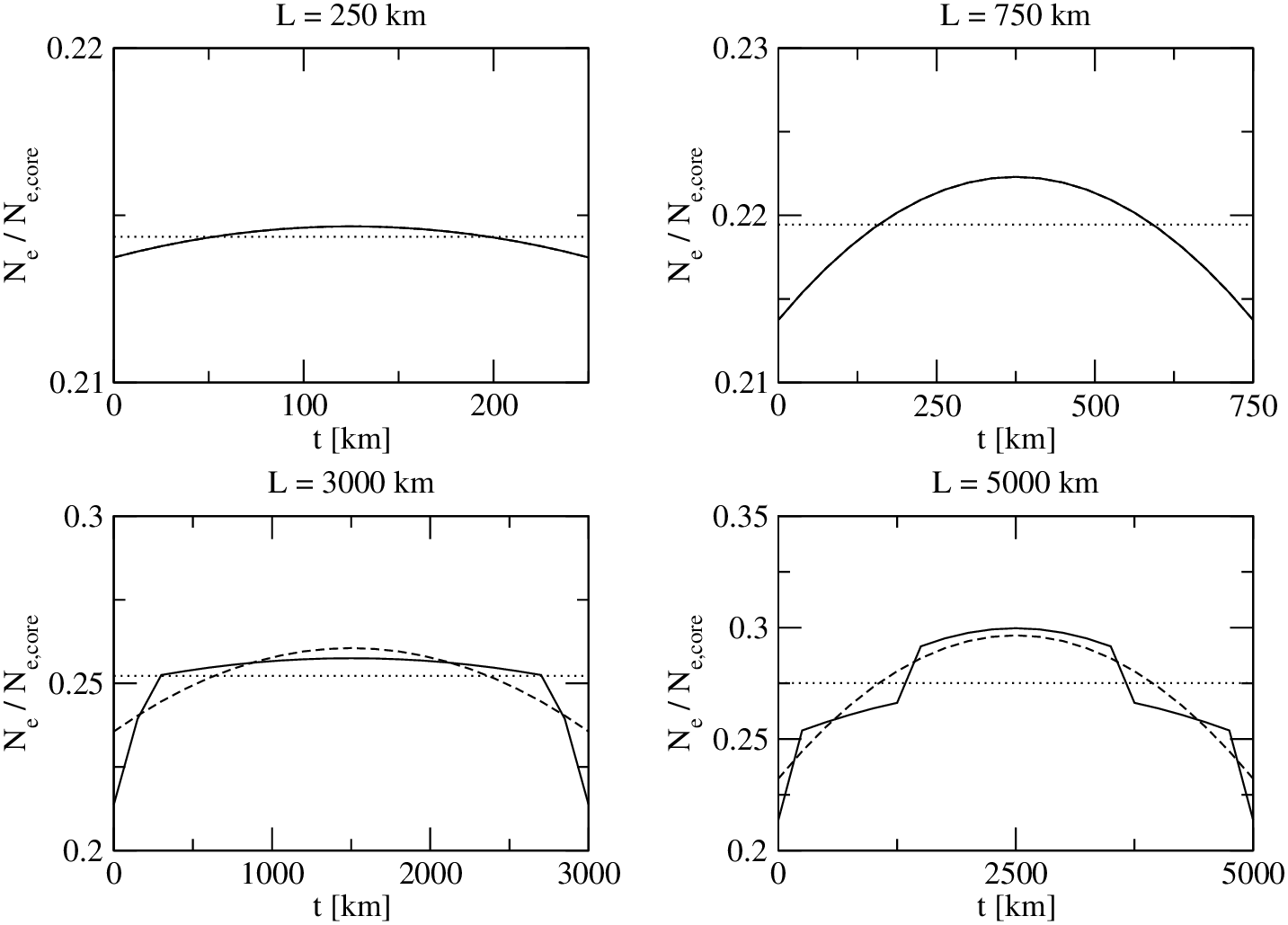}
\caption{The matter density profiles for different baseline lengths
  $L$ according to the PREM. The solid curves are the exact profiles,
  the dashed curves are profiles approximated by a second order
  polynomial, and the dotted lines are the average matter densities,
  \emph{i.e.}, the matter density of approximations using constant
  electron number density. For $L = 250$ km and 750 km the exact
  profiles and the approximations using a second order polynomial are
  practically the same and as a result they are not distinguishable in
  the figure.}
\label{fig:profiles}
\end{center}
\end{figure}

\begin{figure}
\begin{center}
\includegraphics[width=12cm]{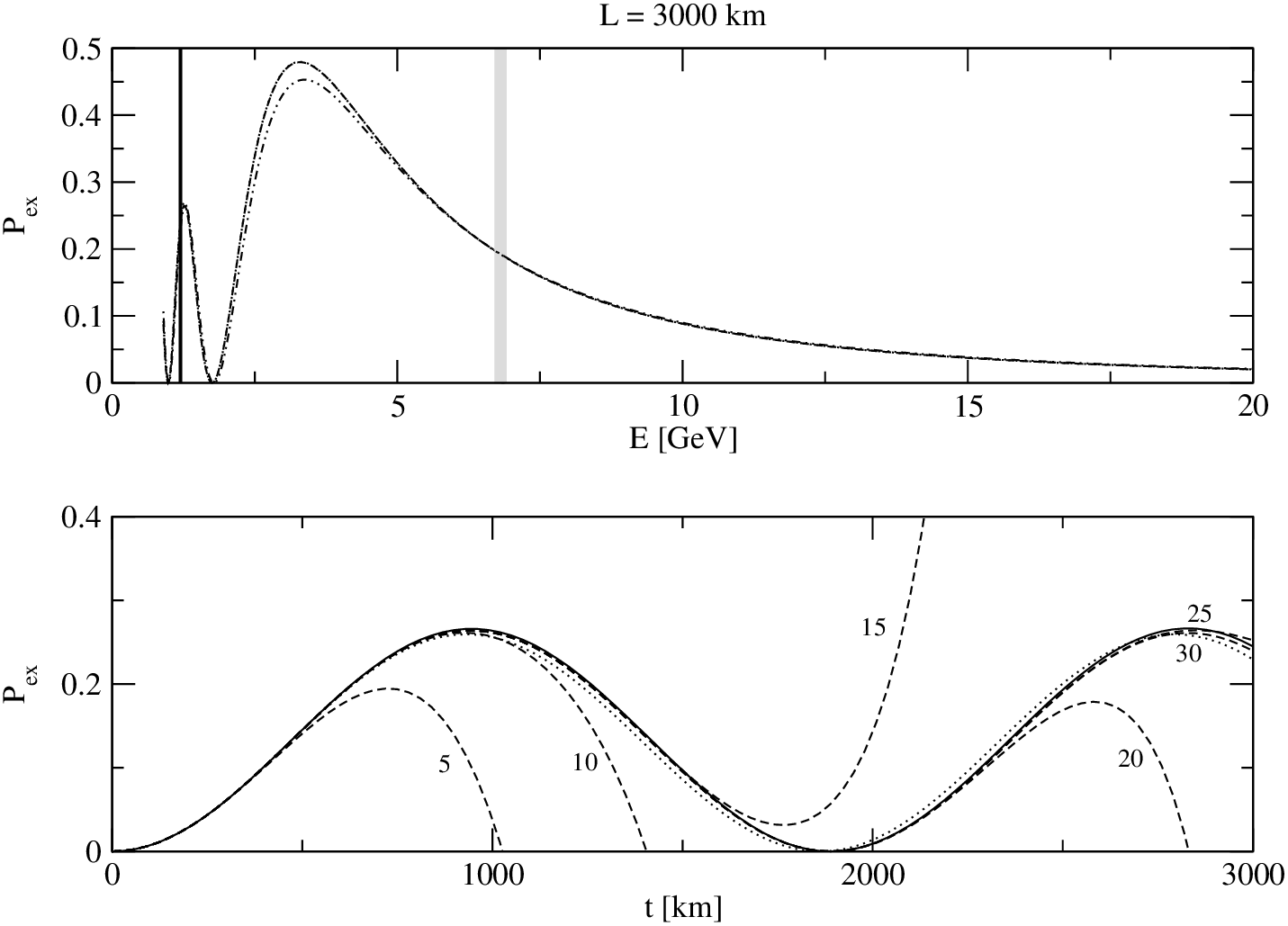}
\caption{The energy spectrum for the neutrino oscillation probability
  $P_{ex}$ using 35 terms of our series solution (upper panel) and the
  convergence of the time evolution for $E = 1.2$ GeV (lower panel).
  Here we assume a baseline of $L = 3000$ km and using the PREM
  profile for the electron number density. In the energy spectrum, the
  dashed curve corresponds to the numerical solution using the PREM
  profile, the dotted curve corresponds to our series solution, and
  the dash-dotted curve corresponds to the approximation using
  constant matter density. The solid vertical line corresponds to the
  energy used in the lower panel and the series solution is not
  plotted in the shaded region where it breaks down numerically. In
  the time evolution plot, the solid curve corresponds to the
  numerical solution, the dotted curve to the solution for the
  constant matter density approximation, the dashed curves correspond
  to our series solution using different numbers of terms, and the
  numbers correspond to the number of terms used for each of these
  curves.}
\label{fig:PREM3000}
\end{center}
\end{figure}

For the PREM electron number density profile, which is the interesting
profile in, for example, long-baseline neutrino oscillation
experiments, we use the expansion in Legendre polynomials and truncate
the series using $N=2$ for definiteness. In effect, this corresponds
to projecting the function $V(t)$, which is an element in the vector
space of real functions on the interval $[0,L]$, onto the subspace of
second order polynomial functions on $[0,L]$, using the inner product
\begin{equation}
\left< f,g\right> = \int_0^L f(x)g(x) {\rm d}x.
\end{equation}
In Fig.~\ref{fig:profiles}, we plot the electron number density
profiles for the baseline lengths $L = 250$ km, 750 km, 3000 km, and
$5000$ km along with the second order polynomial approximations and
the constant approximations.
In Fig.~\ref{fig:PREM3000}, we plot the energy spectra and time
evolution at $E = 1.2$ GeV for a baseline of $L = 3000$ km for the
PREM profile.
Again, our solution is not plotted in the shaded region in which it
breaks down numerically for the same reasons as discussed
previously. In this case, it is apparent that if the number of terms
used in the series expansion is large enough, our solution is a
significant improvement from the constant matter density
approximation.

Clearly, the approximation of using a second order polynomial for the
electron number density gives a very good reproduction of the
numerical solution (which uses the profiles obtained from the
PREM). As can be seen in the time evolution plot, the error made is
barely noticeable until approximately one and a half oscillations,
\emph{i.e.}, for lower neutrino energies if the baseline length $L$ is
kept fixed. This is in good agreement with the results obtained in
Ref.~\cite{Ota:2000hf}, where the effective potential is expanded in a
Fourier series, as well as Ref.~\cite{Ohlsson:2001ck}, which shows that
details of the effective potential that are smaller than the
oscillation length cannot be resolved by neutrino oscillations.  As
noticed in both of the earlier cases, about 20 terms are needed in the
series expansion in order to reproduce one full oscillation.

\begin{figure}
\begin{center}
\includegraphics[width=12cm]{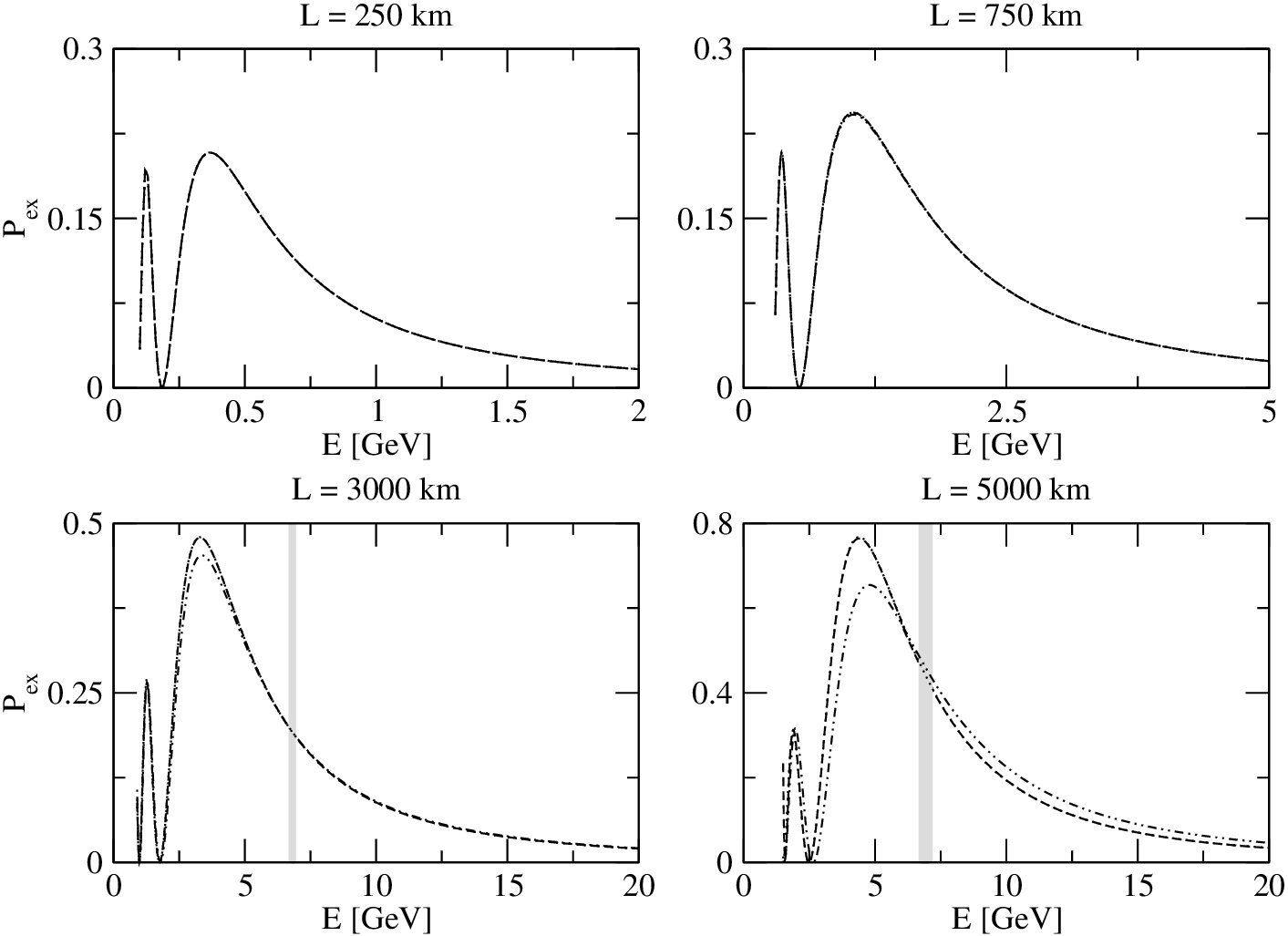}
\caption{The energy spectra of the neutrino oscillation probability
$P_{ex}$ for different baselines using the PREM profile for the
electron number density. Again, we have used 35 terms from our series
solution for each of these energy spectra and the dashed curves
correspond to the numerical solutions, the dotted curves correspond to
our series solution, and the dash-dotted curves correspond to the
constant density approximations. The series solution is not plotted in
the shaded regions where it breaks down numerically.}
\label{fig:PREMES}
\end{center}
\end{figure}

For the PREM profile, we are also interested in a number of other
baseline lengths. In particular, in Fig.~\ref{fig:PREMES}, we plot the
energy spectra for the baseline lengths $L = 250$ km, 750 km, 3000 km,
and $5000$ km.
For the baseline lengths $L = 250$ km and 750 km, there is no
noticeable difference between the numerical solution, our exact
solution, and the approximation using constant electron number
density. This is to be expected as the electron number density does
not vary significantly for these baseline lengths (see
Fig.~\ref{fig:profiles}). However, for both $L = 3000$~km and 5000 km,
we do observe a difference between the constant electron number
density approximations and the other two solutions. Again, we can
conclude that the approximation with a second order polynomial for the
electron number density agrees remarkably well with the numerical
calculation using the profile obtained directly from the PREM. Note
that the comparison of the energy spectra for different matter density
profiles and baselines has been studied before \cite{Freund:1999vc}.

For baselines longer than $L = 5000$ km, the region near the resonance
for $V = V_0$ tends to expand and ruin the numerical convergence of
our solution. Also, this region expands if we include more terms of
the series expansion. For the baseline lengths below $L = 5000$ km,
this can be somewhat compensated by using fewer terms of the series
expansion for high energies to avoid the numerical cancellation
effects and more terms for lower energies to obtain a nice
convergence.

\section{Other methods of solving the neutrino flavor evolution}
\label{sec:methods}

In general, there have been numerous methods on how to solve the
problem of neutrino flavor evolution in matter
\cite{Wolfenstein:1978ue,Ohlsson:1999um,Akhmedov:1988kd,Barger:1980tf,Lehmann:2000ey,Osland:1999et,Smirnov:1987mk,Balantekin:1988aq,Ioannisian:2004jk,Fishbane:2000dc}.
First of all, there is the obvious formal solution using a
time-ordered exponential, \emph{i.e.},
\begin{equation}
\ket{\nu(t)} = T\left[\exp\left(-{\rm i}\int_0^t H(\tau){\rm d}\tau
 \right)\right] \ket{\nu(0)}.
\end{equation}
This solution is exact, but it is not very helpful in actual
calculations due to the nature of the time-ordered exponential. A way
of solving this problem is to use a discretization of the effective
potential \cite{Ohlsson:1999um}\footnote{Similar applications of
neutrino flavor evolution in matter consisting of two density layers
using two flavors have been discussed in
Refs.~\cite{Akhmedov:1988kd}.}. The effective potential is then
divided into a finite number of layers with constant effective
potentials (\emph{i.e.}, constant electron number density), which
approximate a given effective potential. Clearly, when the number of
layers goes to infinity, one regains the time-ordered exponential.

There is also the pure numerical approach to the problem, where the
neutrino flavor evolution is easily solved numerically for any
effective potential. While this gives the possibility of actually
calculating numerical values for the neutrino oscillation
probabilities, it does not offer any insight into how these
probabilities vary with different parameters.

In addition, the neutrino flavor evolution can be analytically solved
for some specific effective potentials. Examples are the constant (two
flavors \cite{Wolfenstein:1978ue} and three flavors
\cite{Barger:1980tf}),
linearly \cite{Lehmann:2000ey} and exponentially \cite{Osland:1999et}
varying effective potentials.

Moreover, a widely used solution is the adiabatic solution
\cite{Smirnov:1987mk,Balantekin:1988aq}, where the effective potential
is assumed to change slowly, so that there are no transitions between
different matter eigenstates of the full Hamiltonian. This
approximation can be derived by, for example, using the
Wentzel--Kramers--Brillouin (WKB) method \cite{Balantekin:1988aq},
where also higher order corrections due to non-adiabatic transitions
can be calculated.

There have also been earlier efforts to write the neutrino evolution
equations as ordinary non-linear differential equations, see for
example Ref.~\cite{Fishbane:2000dc}. However, such equations have
generally been complex differential equations for the probability
amplitudes and not, as in the present case, real differential
equations for the probabilities. Also, the solutions in such cases
have been made for special cases of the effective potential and not as
series solutions valid for all effective potentials.

Lately, approximate solutions, valid when the effective potential $V
\ll \Delta m^2/(2E)$, have been presented
\cite{Ioannisian:2004jk} and applied to oscillations
of solar neutrinos and the solar neutrino day-night effect. However,
these solutions are not valid for the baseline lengths and neutrino
energies which we have treated numerically.

\section{Summary and conclusions}
\label{sec:summary}

We have shown that solving the general problem of two flavor neutrino
oscillations with an arbitrary effective potential is equivalent with
solving the non-linear ordinary differential equation
\begin{equation}
(\ddot p + G p)^2 = F(t)[G(1-p^2)-\dot p^2],
\end{equation}
where $G \equiv [\Delta m^2/(2E)]^2\sin^2(2\theta)$, $F(t) \equiv [\Delta
m^2\cos(2\theta)/(2E) - V(t)]^2$, and the neutrino oscillation probability
$P_{ex}$ is given by
\begin{equation}
P_{ex} = \frac 12 (1-p).
\end{equation}
We have presented an exact solution [see Eqs.~(\ref{eq:pn}) and
(\ref{eq:terms})] to this equation by adopting the method of series
expansion of both the solution and the effective potential $V(t)$ and
demonstrated the numerical convergence of this solution for a number
of different effective potentials. In all cases investigated, about 20
terms in the series expansion are required to reproduce one period of
oscillation. We have also seen that for the neutrino energies and
baselines considered, the energy spectra of the neutrino oscillation
probability is well reproduced by approximating the effective
potential by a second order polynomial.

\begin{acknowledgments}
This work was supported by the Swedish Research Council
(Vetenskapsr{\aa}det), Contract Nos.~621-2001-1611, 621-2002-3577, the
G{\"o}ran Gustafsson Foundation, and the Magnus Bergvall Foundation.
\end{acknowledgments}

\end{document}